\documentclass[10pt]{iopart}
\usepackage{graphicx}
\def\sun{\hbox{$\odot$}}
\def\2F1{~_2F_1}
\usepackage{iopams}
\def\aap{A\&A\,  }
\def\aj{AJ  }
\def\apj{ApJ\,  }
\def\apjl{ApJ\,  }
\def\araa{ARA\&A  }

\def\mnras{MNRAS\,  }

\def\sovast{Soviet Astronomy} 
\begin{document}
\title
{
The Luminosity Function of Galaxies as
modeled by a left truncated 
beta distribution
}
\vskip  1cm
\author     {L. Zaninetti}
\address    {Dipartimento  di Fisica ,
 via P.Giuria 1,\\ I-10125 Turin,Italy }
\ead {zaninetti@ph.unito.it}

\begin{abstract}
A first new luminosity functions of galaxies can be
built starting from
a left truncated beta probability
density function
, which is characterized
by four parameters.
In the astrophysical conversion, the number of parameters
increases by one, due to the addition
of the overall density of galaxies.
A second  new galaxy luminosity function is built
starting from
a left truncated beta probability
for the mass of galaxies once
a simple nonlinear relationship
between mass and luminosity is assumed;
in this case the number of parameters is six
because the overall density of galaxies
and a parameter that regulates mass and luminosity are
added.
The two new galaxy luminosity functions
with finite boundaries were tested
on the Sloan Digital Sky Survey (SDSS)
in five different
bands;
the results  produce a "better fit"
than the Schechter luminosity function in two of
the five bands considered.
A modified  Schechter  luminosity function with four
parameters has been also analyzed. 
\end{abstract}
\vspace{2pc}
\maketitle

\section{Introduction}
The standard  luminosity function for galaxies (LF)
in the last forty years has  been
represented by the
Schechter LF
\begin{equation}
\Phi (L) dL  = (\frac {\Phi^*}{L^*}) (\frac {L}{L^*})^{\alpha}
\exp \bigl ( {-  \frac {L}{L^*}} \bigr ) dL \quad  ,
\label{equation_schechter}
\end {equation}
where $\alpha$ sets the slope for low
values of $L$ , $L^*$
is the
characteristic luminosity and
$\Phi^*$ is the normalization,
see  \cite{schechter}.
This  LF is defined in the
in the interval
$[0, \infty]$  and has replaced other LFs 
presented by 
\cite{Zwicky1957,kiang1961,Abell1965,Arakelyan1970}.

The goodness  of
the fit can be
evaluated  by
the merit function $\chi^2$
\begin{equation}
\chi^2 =
\sum_{j=1}^n ( \frac {LF_{theo} - LF_{astr} } {\sigma_{LF_{astr}}})^2
\quad ,
\label{chisquare}
\end{equation}
where $n$ is number of data and the two
indexes $theo$ and $astr$
stand for theoretical
and astronomical, LF  respectively.
The value  of  $ \chi_S^2 $
which   characterizes the
Schechter LF for  a given astronomical catalog,
i.e. the Sloan Digital Sky Survey (SDSS)
in five different
bands,
represents a standard  value to improve.
Many new LFs has been derived in the last years
and we report some examples.
A
 two-component Schechter-like function is ,
see~\cite{Driver1996}
\begin{eqnarray}
L_{max} > L > L_{Dwarf} : \quad
\Phi (L) dL = (\frac {\Phi^*}{L^*}) (\frac {L}{L^*})^{\alpha}
\exp \bigl ( {- \frac {L}{L^*}} \bigr ) dL \quad,
\nonumber \\
 \\
L_{Dwarf} > L > L_{min} : \quad
\Phi (L) dL = (\frac {\Phi_{Dwarf}}{L^*}) (\frac {L}{L_{Dwarf}})^{\alpha_{
Dwarf}}
 dL \quad,
\nonumber
\end {eqnarray}
where
\begin{eqnarray}
\Phi_{Dwarf} = \Phi^* (\frac {L_{Dwarf} }{L^*})^{\alpha}
\exp \bigl ( {- \frac {L_{Dwarf} }{L^*}} \bigr )
\quad.
\nonumber
\end{eqnarray}
This two-component LF defined between
the maximum luminosity, $L_{max}$,
 and the minimum luminosity, $L_{min}$,
has five parameters because two additional parameters
have been added:
$L_{Dwarf}$ which represents the magnitude where
dwarfs first dominate over giants and ${\alpha_{Dwarf}}$
which regulates
the faint slope parameter for the dwarf population.

In order to
fit the case
of extremely low luminosity galaxies
a  double Schechter LF with five
parameters, see~\cite{Blanton_2005}, was introduced:
\begin{equation}
\Phi(L) dL = \frac{dL}{L_\ast}
 \exp(-L/L_{\ast}) \left[
\phi_{\ast,1}
\left( \frac{L}{L_{\ast}} \right)^{\alpha_1} +
\phi_{\ast,2}
\left( \frac{L}{L_{\ast}} \right)^{\alpha_2}
 \right]
\quad,
\end{equation}
where the parameters $\Phi^*$ and $\alpha$ which characterize
the Schechter LF have been doubled in $\phi_{\ast,1}$
and $\phi_{\ast,2}$.
The strong dependence of LF
on different environments such as voids,
superclusters and supercluster cores was
analyzed by \cite{Einasto_2009} with
\begin{equation}
F (L) \mathrm{d}L \propto (L/L^{*})^\alpha (1 +
(L/L^{*})^\gamma)^{(\delta-\alpha)/\gamma}
\mathrm{d}(L/L^{*}),
\label{eq:abell}
\end{equation}
where $\alpha$ is the exponent
at low luminosities $(L/L^{*}) \ll 1$,
$\delta$ is the exponent at high luminosities
$(L/L^{*}) \gg 1$,
$\gamma$ is a parameter of transition between the
two power laws, and $L^{*}$ is the characteristic luminosity.
Another LF
starts  from the
probability  density function (PDF)
that models area and volumes of the Voronoi Diagrams
and introduces the mass-luminosity relationship,
see  \cite{Zaninetti2008};
in the SDSS  case   $ \chi^2 > \chi_S^2$.
A last  example  is represented
by  three new LFs deduced
in the framework of
generalized gamma PDF, see
\cite{Zaninetti2010f};
in the SDSS case   $ \chi^2 <  \chi_S^2$.
All  the previous LFs cover the range
$[0, \infty]$    and therefore
the analysis of finite upper
and lower boundaries
can be a subject of investigation.
Another interesting observational fact
is that at low values of luminosity
(high absolute
magnitude) the  observed LF has an
approximate  constant  value.
The left truncated beta with scale  PDF
recently  derived  , see  equation (34)
in  \cite{Zaninetti2013a},   satisfies
the two issues previously raised.
In order to explore the  connection between  LF 
and  evolution of galaxies  see 
\cite{vandenBosch2003,Yang2003,Cooray2005a,Cooray2005b,
Cooray2005c,Tinker2005,Tinker2007}.

Here we analyze in Section \ref{lflinear}
a left  truncated beta
LF for galaxies  and  in
Section \ref{lfnonlinear} a
left  truncated beta  for mass  which transforms
itself in a LF for galaxies through  a nonlinear
mass-luminosity  relationship.
Section \ref{secbrazilian}  reports a recent LF for galaxies
which has a finite boundary at the bright end.

\section{A linear mass-luminosity relationship}
\label{lflinear}
A new PDF , the left truncated beta with scale
has been recently derived , see
formula (34) in  \cite{Zaninetti2013a}.
Once the random variable $X$ is substituted
with the luminosity $L$
we obtain  a new  LF  for galaxies,
$\Psi$ ,
\begin{equation}
\Psi (L) dL  =
K_L \, \Psi^*
{L}^{\alpha-1} \left( L_{{b}}-L \right) ^{\beta-1}dL
\quad , \label{betalf}
\end {equation}
where $\Psi^*$  is  a normalization factor which defines the
overall density of galaxies , a  number  per cubic $Mpc$.
The constant $K_L$ is 
\begin{eqnarray}
K_L = \frac{A}{B}               \\
A   = &-\alpha\,\Gamma  \left( \alpha+\beta \right)\nonumber \\
B   = &{L_{{b}}}^{\beta-1}
{\mbox{$_2$F$_1$}(\alpha,-\beta+1;\,1+\alpha;\,{\frac {L_{{a}}}{L_{{b}}}})}
{L_{{a}}}^{\alpha}\Gamma  \left( \alpha+\beta \right)
\nonumber  \\
~&-{L_{{b}}}^{
\beta+\alpha-1}\Gamma  \left( 1+\alpha \right) \Gamma  \left( \beta
 \right)
           \quad ,      \nonumber
\end{eqnarray}
and  $L_a$, $L_b$ are   the lower, upper
values  in luminosity and
${\2F1(a,b;\,c;\,z)}$ is the
regularized hypergeometric
function
\cite{Abramowitz1965,Seggern1992,Thompson1997,Gradshteyn2007,NIST2010}.
The hypergeometric function with real arguments
is implemented in  the FORTRAN subroutine MHYGFX 
in \cite{Zhang1996}.
The averaged luminosity,
$ { \langle L \rangle } $, is:
\begin{eqnarray}
{ \langle L \rangle }
= K_L \, \Psi^*  \frac{A_N}{B_D} \quad ,
\end{eqnarray}
where
\begin{eqnarray}
A_N=&
-{L_{{b}}}^{\beta-1}{L_{{a}}}^{1+\alpha}
{\mbox{$_2$F$_1$}(-\beta+1,1+\alpha;\,2+\alpha;\,{\frac
{L_{{a}}}{L_{{b}}}})}
\Gamma  \left( 1+\alpha+\beta \right) \nonumber \\
~&
 +{L_{{b}}}^{\alpha+\beta}\Gamma
 \left( 2+\alpha \right) \Gamma  \left( \beta \right) \nonumber \\
B_D=&
\left( 1+\alpha \right) \Gamma  \left( 1+\alpha+\beta \right)
\quad . \nonumber
\end{eqnarray}
The relationships connecting the
absolute magnitude $M$,
$M_a$ and  $M_b$
 of a
galaxy to respective  luminosities  are:
\begin{equation}
\frac {L}{L_{\sun}} =
10^{0.4(M_{\sun} - M)}
\; ,
\frac {L_a}{L_{\sun}} =
10^{0.4(M_{\sun} - M_a)}
\;
, \frac {L_b}{L_{\sun}} =
10^{0.4(M_{\sun} - M_b)}
\; ,
\label{magnitudes}
\end {equation}
where $M_{\sun}$ is the absolute magnitude
of the sun in the considered band.
The  LF  in magnitude is
\begin{eqnarray}
\Psi (M) dM   =
-K_M \, \Psi^*
\,0.4\, \left( {10}^{- 0.4\,{\it am}+ 0.4\,{\it M_{\sun}}} \right) ^{
\alpha-1} \times & ~\nonumber \\
\left( {10}^{- 0.4\,{\it  M_b}+ 0.4\,{\it M_{\sun}}}-{10}^{-
 0.4\,{\it am}+ 0.4\,{\it M_{\sun}}} \right) ^{\beta-1}
 \times & ~ \nonumber \\
 {10}^{- 0.4\,{
\it am}+ 0.4\,{\it M_{\sun}}}\ln  \left( 10 \right) dM & ~
\quad ,
\label{lfmagnibeta}
\end{eqnarray}
where
\begin{eqnarray}
K_M   = \frac{A_M}{B_M}  &~               \\
A_M   = \alpha\,\Gamma    ( \alpha+\beta   )&~    \nonumber  \\
B_M   =
- \left( {10}^{- 0.4 {\it M_b}+ 0.4 {\it M_{\sun}}} \right) ^{\beta-1}
\times  &~
\nonumber \\
{\mbox{$_2$F$_1$}(\alpha,-\beta+1; 1+\alpha; {\frac {{10}^{- 0.4 {\it
M_a}+ 0.4 {\it M_{\sun}}}}{{10}^{- 0.4 {\it M_b}+ 0.4 {\it M_{\sun}}}}})}
\times  \nonumber &~\\
 \left( {10}^{- 0.4 {\it M_a}+ 0.4 {\it M_{\sun}}} \right) ^{\alpha}
\Gamma  \left( \alpha+\beta \right)
\nonumber &~ \\
 + \left( {10}^{- 0.4 {\it M_b}+
 0.4 {\it M_{\sun}}} \right) ^{\beta+\alpha-1}\Gamma  \left( 1+\alpha
 \right) \Gamma  \left( \beta \right)&~
\quad.
\nonumber
\end{eqnarray}

This data-oriented LF  contains the five parameters
$\alpha$,
$\beta$,
$M_a$,
$M_b$,
$\Psi^*$ which can be derived from the
operation of fitting
the observational data
and  $M_{\sun}$  which  characterize the considered band.
The  number of variables  can be reduced to three
once $M_a$ and $M_b$ are identified with
the maximum and the minimum   absolute magnitude of  the
considered sample.
The numerical analysis points
toward value  of
$\alpha \approx  0$.
Once as an example
we fix $\alpha=0.01$ we  have only to  find
the two parameters $\beta$ and  $\Psi^*$.
The test on the new LF  were performed on the data
of the  Sloan Digital Sky Survey (SDSS)
which has  five bands
$u^*$,
$g^*$ ,
$r^*$ ,
$i^*$ ,
and
$z^*$ , see  \cite{Blanton_2003}.
Table~\ref{databetalf}
reports
the  bolometric magnitude,
the numerical values obtained for the new LF ,
the number , $N$, of the astronomical sample ,
the $\chi^2$  of the new LF  ,
the reduced  $\chi^2$ , $ \chi_{red}^2$,  of the new LF  ,
the $\chi_S^2$  of the  Schechter LF    ,
the reduced  $\chi^2$ ,$ \chi_{S,red}^2$, of Schechter LF  and
the $\chi_S^2$ computed as in \cite{Blanton_2003}
for the five bands here  considered.

 \begin{table}
 \caption[]{
 Parameters of fits
 in SDSS Galaxies of
 LF as  represented by
 formula (\ref{lfmagnibeta}) in which
 the number of free parameters is 5.
}
 \label{databetalf}
 \[
 \begin{array}{lccccc}
 \hline
parameter    &  u^*   &  g^* & r^*  & i^* & z^*  \\ \noalign{\smallskip}
 \hline
 \noalign{\smallskip}
 M_{\sun}
 &  6.38
 &  5.07
 &  4.62
 &  4.52
 &  4.48
 \\
M_{a}
 & -15.90
 & -16.56
 & -16.81
 & -17.27
 & -17.56
 \\
M_{b}
 & -21.48
 & -22.95
 & -24.03
 & -24.45
 & -24.75
 \\

\Psi^* [h^3~Mpc^{-3}]
 & 0.040
 & 0.044
 & 0.042
 & 0.038
 & 0.038
 \\
\alpha
 & 0.01
 & 0.01
 & 0.01
 & 0.01
 & 0.01
 \\
\beta
 & 23.44
 & 23.83
 & 27.88
 & 27.56
 & 27.36
 \\
N & 483
 & 599
 & 674
 & 709
 & 740
 \\

\chi^2
 & 570
 & 1032
 & 4342
 & 3613
 & 6252
 \\
\chi_{red}^2
 & 1.19
 & 1.73
 & 6.49
 & 5.13
 & 8.5
 \\

\chi_S^2
 & 330
 & 753
 & 2260
 & 2282
 & 3245
\\
\chi_{S,red}^2
 & 0.689
 & 1.263
 & 3.368
 & 3.232
 & 4.403
\\

\chi^2 -Blanton~2003
 & 341
 & 756
 & 2276
 & 2283
 & 3262
\\
 \end{array}
 \]
 \end {table}

\section {A non linear mass-luminosity relationship}
\label{lfnonlinear}
We assume that the mass of the galaxies
,${\mathcal{M}}$,
is distributed with a left truncated  PDF
as represented
by formula (34) in  \cite{Zaninetti2013a}.
The distribution for the mass is
$
\Psi
({\mathcal{M}};{\mathcal{M}}_a,{\mathcal{M}}_b ,\alpha,\beta)
$
where
${\mathcal{M}}_a$ is the lower  mass and
${\mathcal{M}}_b$ is the upper  mass.
The mass  is supposed to scale  as
\begin{equation}
{\mathcal {M}} = L ^{\frac{1}{c}}
\quad  ,
\end{equation}
where  $c$ is the parameter  that  regulates
the non linear mass-luminosity relationship.
The  differential of the mass  is
\begin{equation}
d {\mathcal {M}} = \frac{L ^{\frac{1}{c} -1}}{c} dL
\quad  .
\end{equation}
As a   consequence  the  LF representing the
non linear mass-luminosity relationship
is
\begin{equation}
\Psi_{ML}  (L) dL  =
K_{ML} \, \Psi^*_{ML}
\frac
{
\left( {L}^{{c}^{-1}} \right) ^{\alpha-1} \left( {L_{{b}}}^{{c}^{-1}}
-{L}^{{c}^{-1}} \right) ^{\beta-1}{L}^{{c}^{-1}}
}
{
cL
}
dL
\quad  ,
\end {equation}
where
\begin{eqnarray}
K_{ML} = \frac{A_{ML}}{B_{ML}}               \\
A_{ML}   =&-\alpha\,\Gamma  \left( \alpha+\beta \right)   \nonumber \\
B_{ML}   =&( {L_{{b}}}^{{c}^{-1}}   ) ^{\beta-1}
{\mbox{$_2$F$_1$}(\alpha,-\beta+1;\,1+\alpha;\,{\frac
{{L_{{a}}}^{{c}^{-1}}}{{L_{{b}}}^{{c}^{-1}}}})}
   ( {L_{{a}}}^{{c}^{-1}}   ) ^{\alpha}\Gamma    ( \alpha+
\beta   ) \nonumber \\
& - ( {L_{{b}}}^{{c}^{-1}}   ) ^{\beta-1+\alpha}
\Gamma    ( 1+\alpha   ) \Gamma    ( \beta   )
 \nonumber   \quad .  \\
\end{eqnarray}
The averaged luminosity, $ { \langle L \rangle } $,
for the non linear mass-luminosity relationship
is:
\begin{eqnarray}
{ \langle L \rangle }
= K_{ML} \, \Psi^*  \frac{A_{NM}}{B_{DM}} \quad ,
\end{eqnarray}
where
\begin{eqnarray}
A_{NM}= {L_{{b}}}^{{\frac {c+\alpha+\beta-1}{c}}}\Gamma  \left( c+\alpha
 \right) \Gamma  \left( \beta \right) c+{L_{{b}}}^{{\frac {c+\alpha+
\beta-1}{c}}}\Gamma  \left( c+\alpha \right) \Gamma  \left( \beta
 \right) \alpha  &~ \nonumber \\
 -{L_{{b}}}^{{\frac {\beta-1}{c}}}
{\mbox{$_2$F$_1$}(-\beta+1,c+\alpha;\,1+\alpha+c;\,{L_{{a}}}^{{c}^{-1}}{L_{{b}}}^{-{c}^{-1}})}
\times  &~   \nonumber \\
{L_{{a}}}^{{\frac {c+\alpha}{c}}}\Gamma  \left( c+\alpha+\beta
 \right)                       & ~  \nonumber \\
B_{DM}= \Gamma  \left( c+\alpha+\beta \right)  \left( c+\alpha \right)
 & \quad .\nonumber
\end{eqnarray}

The  LF  in magnitude for the non linear
mass-luminosity relationship
is
\begin{eqnarray}
\Psi_{ML} (M) dM   =&   \nonumber  \\
-K_{MML} \, \Psi^* [
0.4\,  (   ( {10}^{- 0.4\,{\it am}+ 0.4\,{\it M_{\sun}}}
  ) ^{{c}^{-1}}  ) ^{\alpha-1}  (   ( {10}^{- 0.4\,{
\it M_b}+ 0.4\,{\it M_{\sun}}}  ) ^{{c}^{-1}} 
&  ~
\nonumber  \\
 -( {10}^{- 0.4\,
{\it am}+ 0.4\,{\it M_{\sun}}}  ) ^{{c}^{-1}}  ) ^{\beta-1}
  ( {10}^{- 0.4\,{\it am}+ 0.4\,{\it M_{\sun}}}  ) ^{{c}^{-1}}
\ln   ( 10  )]/c  \quad ,
& ~
\label{lfmagnibetaml}
\end{eqnarray}
where
\begin{eqnarray}
K_{MML}   = &\frac{A_{MML}}{B_{MML}}               \\
A_{MML}   = & \alpha\,\Gamma   ( \alpha+\beta  )    \nonumber  \\
B_{MML}   = & - (   ( {10}^{- 0.4\,{\it M_b}+ 0.4\,{\it M_{\sun}}}
 ) ^{{
c}^{-1}}  ) ^{\beta-1}
\times  \nonumber  \\
~&{\mbox{$_2$F$_1$}(\alpha,-\beta+1;\,1+\alpha;\,{\frac {  ( {10}^{-
0.4\,{\it M_a}+ 0.4\,{\it M_{\sun}}}  ) ^{{c}^{-1}}}{  ( {10}^{-
0.4\,{\it M_b}+ 0.4\,{\it M_{\sun}}}  ) ^{{c}^{-1}}}})} \times \nonumber \\
~& 
  (   ( {10}^{- 0.4\,{\it M_a}+ 0.4\,{\it M_{\sun}}}  ) ^{{
c}^{-1}}  ) ^{\alpha}\Gamma   ( \alpha+\beta  )
\nonumber  \\
~&+
  (   ( {10}^{- 0.4\,{\it M_b}+ 0.4\,{\it M_{\sun}}}  ) ^{{
c}^{-1}}  ) ^{\beta-1+\alpha}\Gamma   ( 1+\alpha  )
\Gamma   ( \beta  )
                    \quad . \nonumber
\end{eqnarray}
Table~\ref{databetalfml}
reports
the numerical values obtained for
the new nonlinear LF
as represented by eqn.
(\ref{lfmagnibetaml}),
the $\chi^2$  of the new LF  and
the $ \chi_{red}^2$   of the new LF
in  the five bands here  considered.

\begin{figure}
 \centering
\includegraphics[width=6cm]{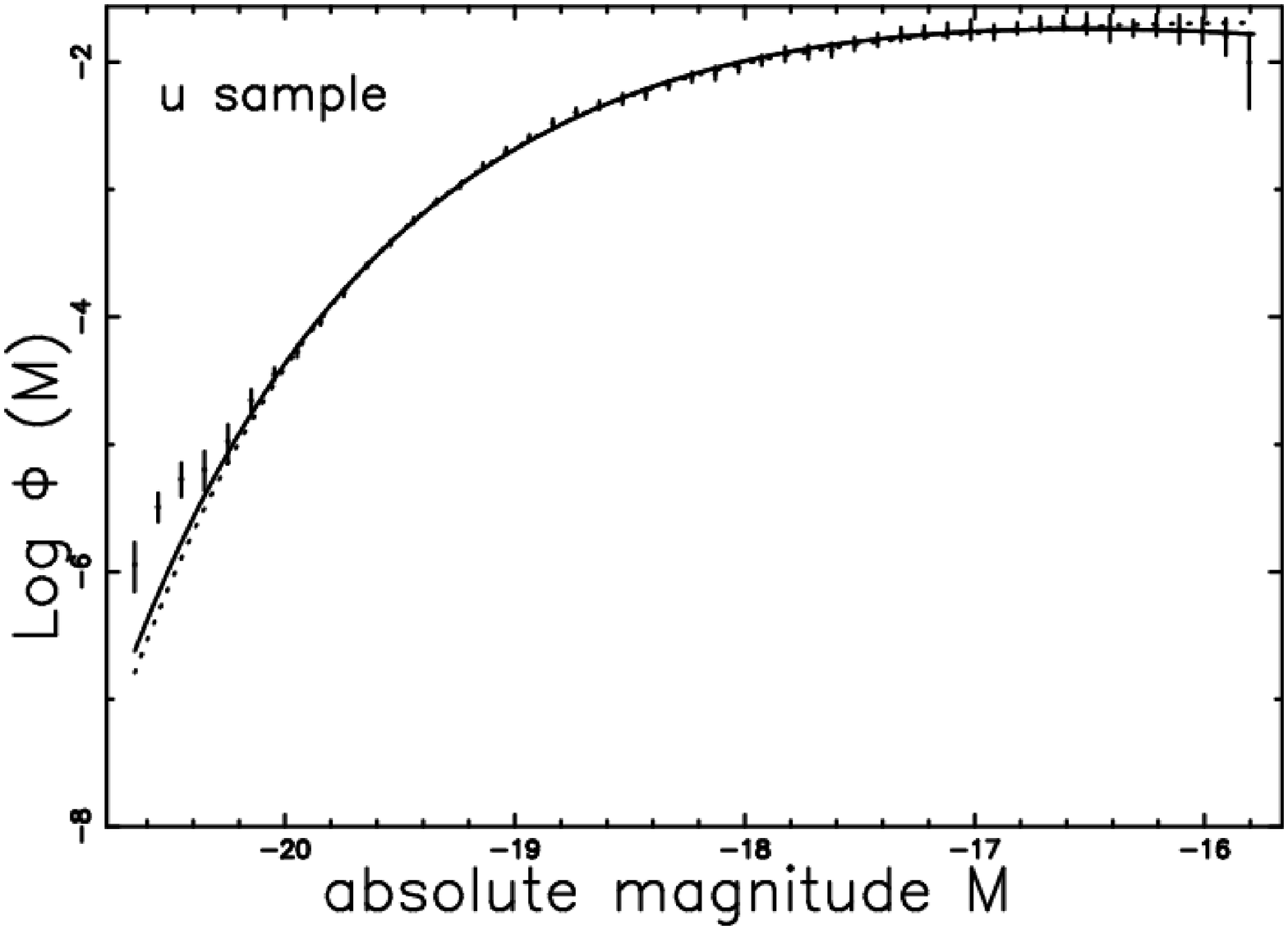}
\caption {The LF data of
SDSS($u^*$) are represented with error bars.
The continuous line fit represents our non linear LF
(\ref{lfmagnibetaml})
and the dotted
line represents the Schechter LF.
 }
 \label{due_u}
 \end{figure}

\begin{figure}
 \centering
\includegraphics[width=6cm]{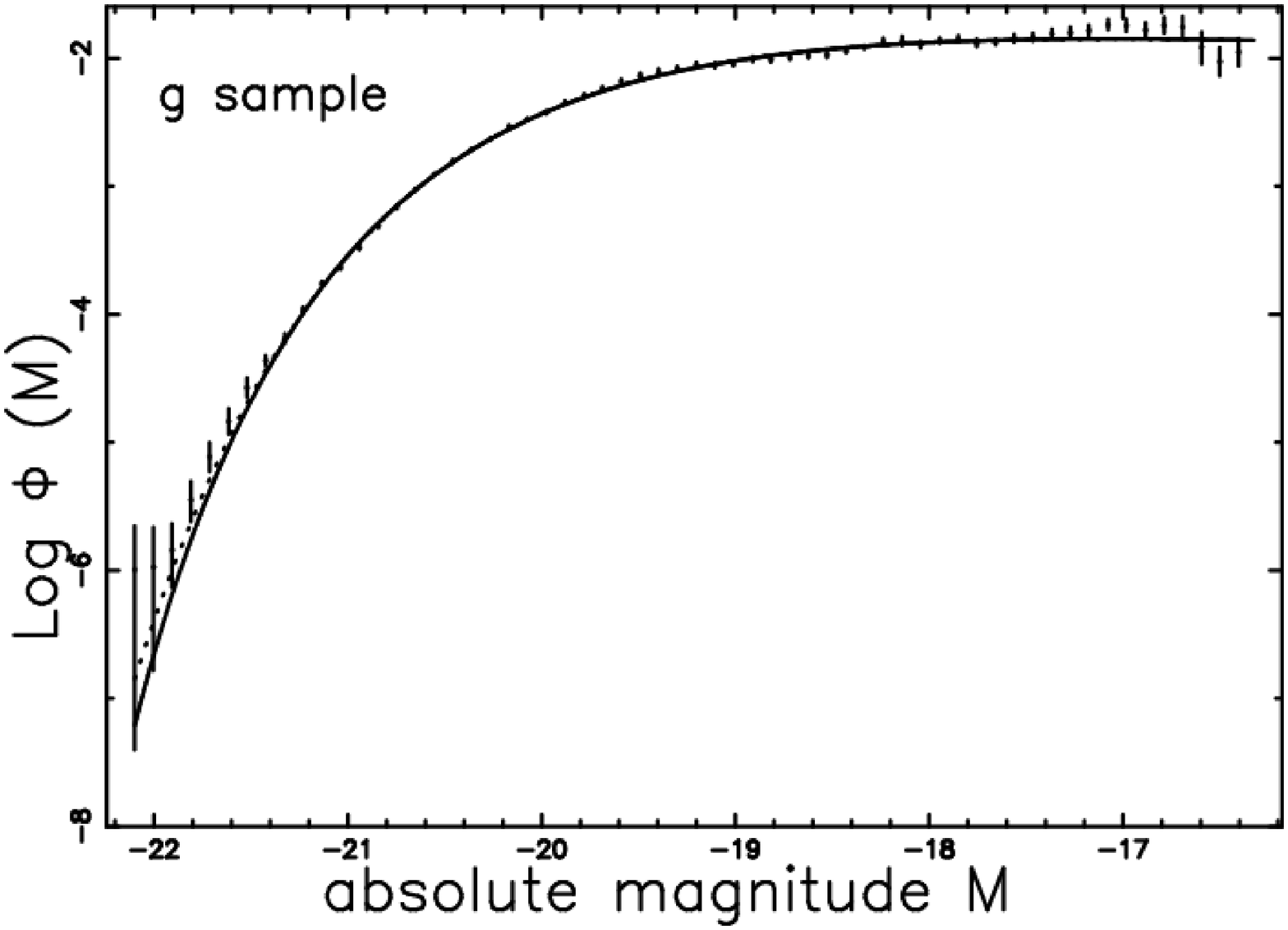}
\caption {The luminosity function data of
SDSS($g^*$) are represented with error bars.
The continuous line fit represents our non linear LF
(\ref{lfmagnibeta})
and the dotted
line represents the Schechter LF.
 }
 \label{due_g}
 \end{figure}

\begin{figure}
 \centering
\includegraphics[width=6cm]{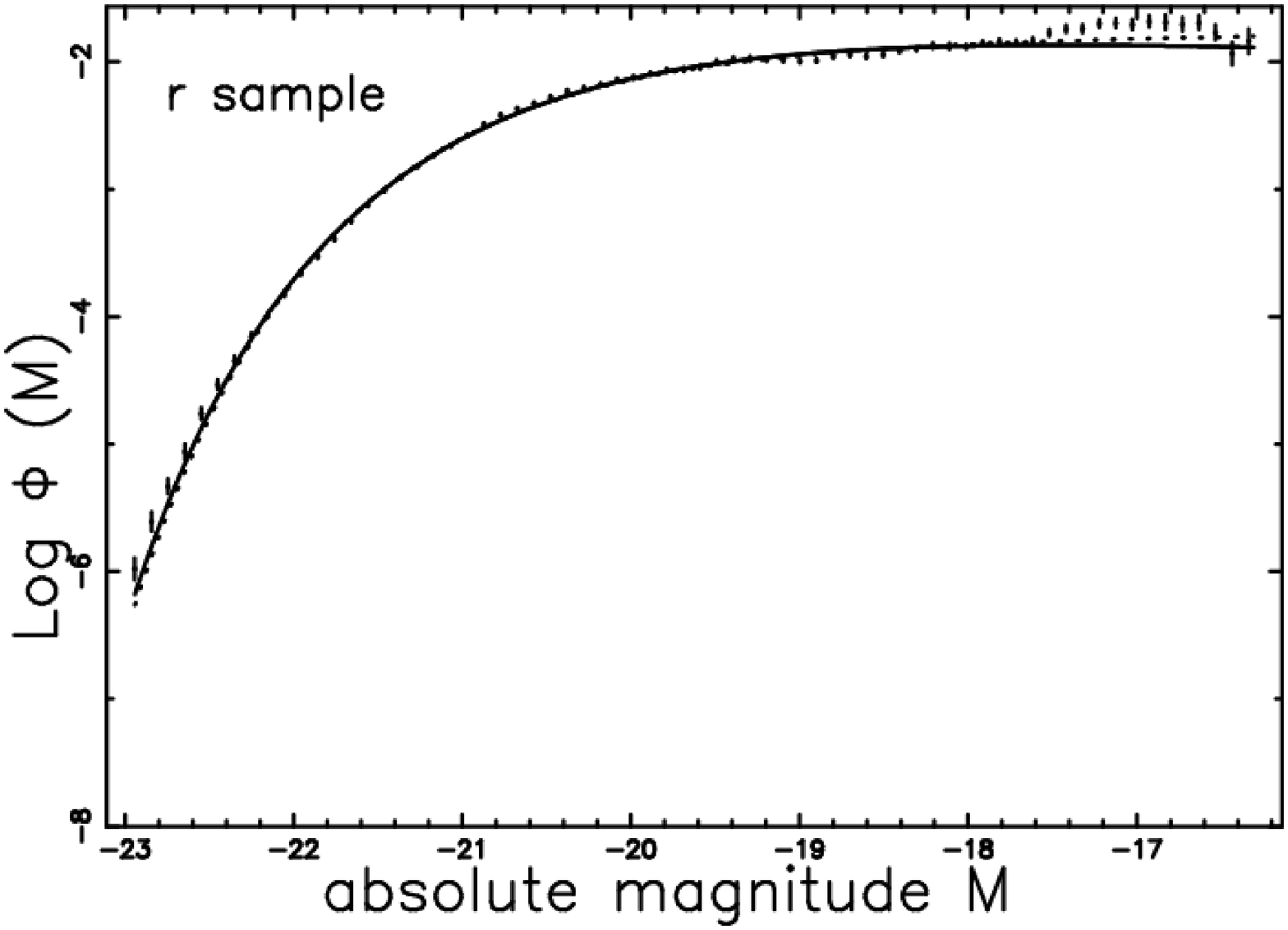}
\caption {The luminosity function data of
SDSS($r^*$) are represented with error bars.
The continuous line fit represents our non linear LF
(\ref{lfmagnibeta})
and the dotted
line represents the Schechter LF.
 }
 \label{due_r}
 \end{figure}

\begin{figure}
 \centering
\includegraphics[width=6cm]{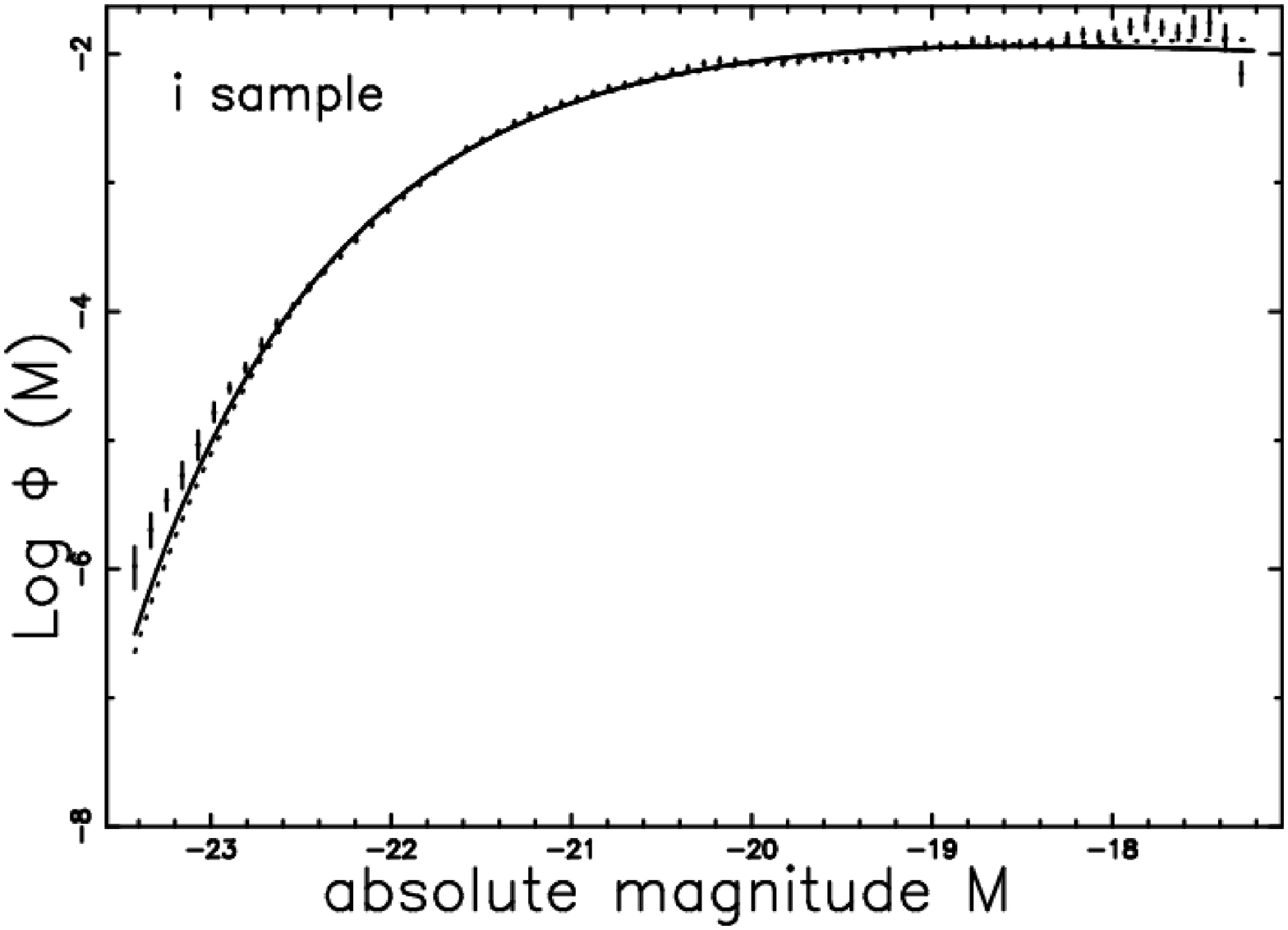}
\caption {The luminosity function data of
SDSS($i^*$) are represented with error bars.
The continuous line fit represents our non linear LF
(\ref{lfmagnibeta})
and the dotted
line represents the Schechter LF.
 }
 \label{due_i}
 \end{figure}

\begin{figure}
 \centering
\includegraphics[width=6cm]{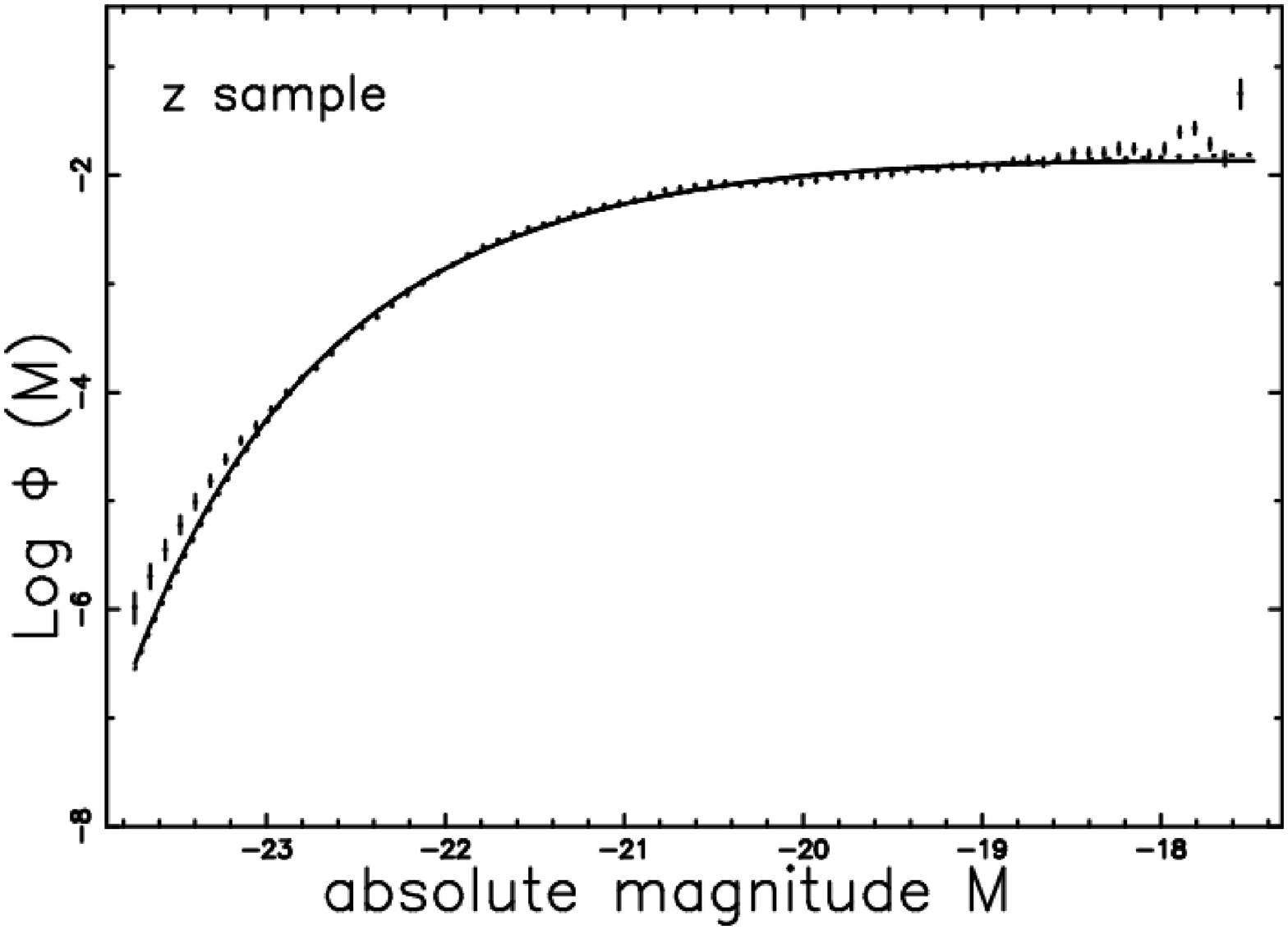}
\caption {The luminosity function data of
SDSS($z^*$) are represented with error bars.
The continuous line fit represents our non linear LF
(\ref{lfmagnibeta})
and the dotted
line represents the Schechter LF.
 }
 \label{due_z}
 \end{figure}

 \begin{table}
 \caption[]{
 Parameters of fits
 in SDSS Galaxies of
 LF as  represented by
our non linear LF
(\ref{lfmagnibetaml})
 formula (\ref{lfmagnibeta}) in which
the number of free parameters is 6.
}
 \label{databetalfml}
 \[
 \begin{array}{lccccc}
 \hline
parameter    &  u^*   &  g^* & r^*  & i^* & z^*  \\ \noalign{\smallskip}
 \hline
 \noalign{\smallskip}
M_{a}
 & -15.38
 & -16.2
 & -16.59
 & -16.99
 & -17.41
 \\
M_{b}
 & -22.3
 & -23.31
 & -24.68
 & -25.02
 & -25.34
 \\

\Psi^* [h^3~Mpc^{-3}]
 & 0.048
 & 0.045
 & 0.046
 & 0.041
 & 0.043
 \\
\alpha
 & 1.0
 & 0.16
 & 0.16
 & 0.3
 & 0.09
 \\
\beta
 & 33.05
 & 30.16
 & 33.92
 & 31.02
 & 32.41
 \\
 c
 & 1.52
 & 1.1
 & 1.25
 & 1.31
 & 1.23
 \\

\chi^2
 & 314
 & 866
 & 2568
 & 2174
 & 3325
 \\
\chi_{red}^2
 & 0.66
 & 1.46
 & 3.84
 & 3.09
 & 4.53
 \\
 \end{array}
 \]
 \end {table}
The Schechter LF, the new LF
represented by formula~(\ref{lfmagnibetaml})
and the data of
SDSS are
reported in
Figure~\ref{due_u},
Figure~\ref{due_g},
Figure~\ref{due_r},
Figure~\ref{due_i} and
Figure~\ref{due_z},
where bands
$u^*$,
$g^*$,
$r^*$,
$i^*$ and
$z^*$
are considered.

A discussion on the distribution of $M/L$ 
in all photometric bands can be found  in
\cite{Kauffmann2003}.
\section{A modified Schechter  LF}
\label{secbrazilian}
An example of  modified Schechter  LF
is represented
by the following
function
\begin{equation}
\Phi_B (L) dL  =
\frac
{
{\it \Phi^*}\, \left( {\frac {L}{{\it L^*}}} \right) ^{\alpha}
 \left( 1-{\frac { \left( \eta-1 \right) L}{{\it L^*}}} \right)
^{\frac{1}{ \eta-1 }}
}
{
{\it L^*}
}
dL
\quad  ,
\end{equation}
where $\eta$ is a new parameter, see \cite{Alcaniz2004}.
This new LF in the case  of
$\eta < 1$ is defined in the range
$0 < L < L_{max} $  where $L_{max}=\frac{L^*}{\eta-1} $
and therefore has a natural upper boundary  which is
not infinity.
In the limit $\lim_{\eta \to 1} \Phi_B (L)$ the
Schechter LF is obtained.
In the case of $\eta > 1$ the average value is
\begin{equation}
{ \langle L \rangle }
=
\frac
{{\it \Phi^*}{\it L^*}\,
\Gamma  \left( 3+\alpha \right) \Gamma  \left( 1+ \left( \eta-1
 \right) ^{-1} \right)  \left( \eta-1 \right) ^{-\alpha-2
}
}
{
\Gamma  \left( 3+\alpha+ \left( \eta-1 \right) ^{-1} \right)  \left(
\alpha+2 \right)
}
\quad ,
\end{equation}
and in the case of $\eta <1 $ the average value is
\begin{equation}
{ \langle L \rangle }
=
\frac
{
{\it \Phi^*}{\it L^*}\,
\left( -\eta+1 \right) ^{-2-\alpha}\,\Gamma
 \left( -{\frac {-1+2\,\eta+\alpha\,\eta-\alpha}{\eta-1}} \right)
\Gamma  \left( 2+\alpha \right)
}
{
\Gamma  \left( - \left( \eta-1 \right) ^{-1} \right)
}
\quad .
\end{equation}

\begin{figure}
 \centering
\includegraphics[width=6cm]{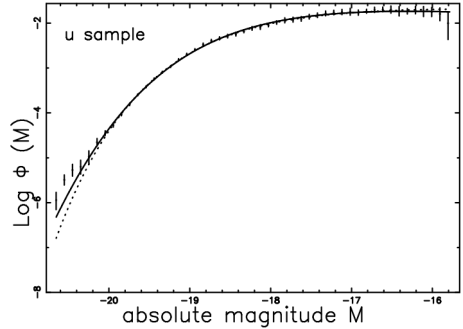}
\caption {The luminosity function data of
SDSS($u^*$) are represented with error bars.
The continuous line fit represents the modified Schechter LF
(\ref{lfbrazilian})
and the dotted
line represents the Schechter LF .
 }
 \label{brazilian_due_u}
 \end{figure}

The distribution  in magnitude is :

\begin{equation}
\Phi(M)dM=
 0.921
 \Phi^*
10^{0.4(M^*-M) (\alpha+1)}
(1- (\eta -1) 10^{0.4(M^*-M)} )
^{\frac{1}{ \eta-1 }}
dM
\quad  .
\label{lfbrazilian}
\end{equation}
As an example the Schechter LF, the modified Schechter LF
represented by formula~(\ref{lfbrazilian})
and the data of
SDSS are
reported in
Figure~\ref{brazilian_due_u} for the $u^*$ band
and Table \ref{databrazilian} reports the results of the fits
in all the five bands of SDSS.

 \begin{table}
 \caption[]{
 Parameters of fits
 in SDSS Galaxies of
the modified Schechter  LF as  represented by
 formula (\ref{lfbrazilian}) in which
 the number of free parameters is 4.
}
 \label{databrazilian}
 \[
 \begin{array}{lccccc}
 \hline
parameter    &  u^*   &  g^* & r^*  & i^* & z^*  \\ \noalign{\smallskip}
 \hline
 \noalign{\smallskip}
M^*
 &  -17.68
 & -19.36
 & -20.30
 & -20.54
 & -20.92
 \\

\Phi^* [h^3~Mpc^{-3}]
 & 0.037
 & 0.022
 & 0.016
 & 0.018
 & 0.017
 \\
\alpha
 & -0.721
 & -0.874
 &  -0.983
 & -0.856
 & -0.942
 \\
\eta
 & 0.959
 & 0.994
 &  0.966
 &  0.938
 &  0.94
 \\
N & 483
 & 599
 & 674
 & 709
 & 740
 \\

\chi^2
 & 252
 & 749
 & 1944
 & 1498
 & 2409
 \\
\chi_{red}^2
 & 0.526
 & 1.25
 &  2.9
 & 2.12
 & 3.27
 \\

\chi_S^2
 & 330
 & 753
 & 2260
 & 2282
 & 3245
\\
\chi_{S,red}^2
 & 0.689
 & 1.263
 & 3.368
 & 3.232
 & 4.403
\\

\chi^2 -Blanton~2003
 & 341
 & 756
 & 2276
 & 2283
 & 3262
\\
 \end{array}
 \]
 \end {table}

\section{Conclusions}
{\bf Motivations}
The observational  fact that the LF for galaxies spans
from a minimum to a maximum value
in absolute magnitude makes attractive the
exploration  of the left truncated  beta LF.
We derived two LFs adopting
the framework of the linear M-L relationship
, see  eqn. (\ref{lfmagnibeta}) ,
and the framework
of the non linear M-L  relationship
, see  eqn. (\ref{lfmagnibetaml}).

{\bf Goodness of fit tests}

We have computed  $\chi^2$ and
$ \chi_{red}^2 $   for  two  LFs  here
derived and compared the  results
with the   Schechter LF,
see Tables \ref{databetalf} and
\ref{databetalfml}.
In the  linear  M-L case
$\chi^2$ and
$ \chi_{red}^2 $  are greater of that of the
Schechter LF.
In the  non linear  M-L case
$\chi^2$ and
$ \chi_{red}^2 $
,  $u^*$ and $i^*$ ,
are  smaller
in two bands over five
in respect to those
of the
Schechter LF.
The  non linear case suggests
a power law behavior with an exponent  $1.1 < c < 1.52$
and therefore M-L relationship has an exponent
greater than 1 but smaller than 4 ,
the theoretical value of the stars.

{\bf Modified Schechter LF}

The modified Schechter LF when $\eta <1$ , which is the case 
of SDSS , has by definition an upper boundary,
see \ref{secbrazilian}.
The  
$\chi^2$ and
$ \chi_{red}^2 $
are  smaller
in five bands over five
in respect to those
of the
Schechter LF.
This goodness of fit is due to the  flexibility introduced  by 
the fourth parameter $\eta$ as outlined  in
\cite{Alcaniz2004}.

{\bf Physical  Motivations}

The Schechter LF is motivated by the Press-Schechter  formalism
on the self-similar  gravitational  condensation.
Here conversely we limited ourself 
to fix a lower and an upper  bound  on the mass of a galaxy
and we translated this requirement   in a PDF for mass
and then  in LF.
Finite-size effects  on the luminosity of SDSS galaxies 
has also been explored in \cite{Taghizadeh-Popp2012}. 

\providecommand{\newblock}{}

\end{document}